\documentclass[aip,jap,reprint,superscriptaddress,footinbib]{revtex4-1}

\usepackage{graphicx}
\usepackage{dcolumn}

\begin{document}

\title{Microwave-induced excess quasiparticles in superconducting resonators measured through correlated conductivity fluctuations} 

\author{P.J. de Visser}
\email{p.j.devisser@tudelft.nl}

\affiliation{SRON National Institute for Space Research, Sorbonnelaan 2, 3584 CA Utrecht, The Netherlands}
\affiliation{Kavli Institute of NanoScience, Faculty of Applied Sciences, Delft University of Technology, Lorentzweg 1, 2628 CJ Delft, The Netherlands}

\author{J.J.A. Baselmans}
\author{S.J.C. Yates}
\author{P. Diener}
\affiliation{SRON National Institute for Space Research, Sorbonnelaan 2, 3584 CA Utrecht, The Netherlands}

\author{A. Endo}
\author{T.M. Klapwijk}
\affiliation{Kavli Institute of NanoScience, Faculty of Applied Sciences, Delft University of Technology, Lorentzweg 1, 2628 CJ Delft, The Netherlands}

\date{\today}

\begin{abstract}

We have measured the number of quasiparticles and their lifetime in aluminium superconducting microwave resonators. The number of excess quasiparticles below 160 mK decreases from 72 to 17 $\mu$m$^{-3}$ with a 6 dB decrease of the microwave power. The quasiparticle lifetime increases accordingly from 1.4 to 3.5 ms. These properties of the superconductor were measured through the spectrum of correlated fluctuations in the quasiparticle system and condensate of the superconductor, which show up in the resonator amplitude and phase respectively. Because uncorrelated noise sources vanish, fluctuations in the superconductor can be studied with a sensitivity close to the vacuum noise.

\end{abstract}

\maketitle
%intro on correlations in qp and condensate
The promise of a long quasiparticle lifetime and a long coherence time makes superconducting circuits popular for use in radiation detection and quantum computation. At low temperature the number of quasiparticles in a superconductor should decrease exponentially. Excess quasiparticles were recently suggested to limit the coherence time of superconducting qubits\cite{jmartinis2009,*gcatelani2011,*hpaik2011,*mlenander2011,*rbarends2011,*acorcoles2011} and the tunnelling rate in single-electron transistors\cite{osaira2012}. Recently, the quasiparticle lifetime in a high-quality aluminium superconducting resonator\cite{rbarends2008c} was shown to be consistent with an excess quasiparticle population inferred from noise measurements\cite{pdevisser2011}. There is a vivid debate on the question of the origin of those excess quasiparticles\cite{jmartinis2009,*gcatelani2011,*hpaik2011,*mlenander2011,*rbarends2011,*acorcoles2011,osaira2012,pdevisser2011}, which mainly focuses on reducing the influence of the environment on the devices under study. Here we show for superconducting aluminium resonators that the environment is well enough under control in our experimental setup to reveal a new source of quasiparticles, namely the microwave readout power of these devices. We show that the saturation in the number of quasiparticles at low temperature (100-150 mK), as inferred from noise measurements, decreases from 72 to 17 $\mu$m$^{-3}$ with a 6 dB decrease of the microwave power. The quasiparticle lifetime increases accordingly from 1.4 to 3.5 ms. 

%microwave resonators and measurement of correlated spectra, here we show...
Microwave resonators are popular devices in radiation detection \cite{pday2003} and circuit quantum electrodynamics \cite{awallraff2004}. The two quadratures of the microwave field in such a resonator are proportional to the real and imaginary part of the conductivity of the superconductor $\sigma_1-i\sigma_2$. The real part corresponds to dissipation in the quasiparticle system and the imaginary part to the kinetic inductance of the condensate \cite{dmattis1958}. We have shown recently that the real part of the conductivity shows quasiparticle number fluctuations \cite{pdevisser2011}. When two quasiparticles recombine, a Cooper pair is formed and when a Cooper pair is broken it leaves two quasiparticles. Therefore, the superconducting condensate fluctuates as well, and one would expect to see these fluctuations in the reactive response of the microwave resonator. However, this reactive response is obscured by the response of two-level fluctuators in the dielectrics surrounding the resonator \cite{jgao2007}. Therefore, we study here the correlation between the dissipative and reactive part of the conductivity in an aluminium resonator. We observe correlated fluctuations in the dissipative and reactive parts of the response, which proves the correlated nature of fluctuations in the quasiparticle system and the condensate. The correlation results in a measurement of fluctuations in the superconductor down to the vacuum noise level, even with a conventional amplifier. The number of quasiparticles and their lifetime are extracted from the fluctuation spectra. 

%Theory
In thermal equilibrium, the average number of quasiparticles per unit volume in a superconductor follows an exponential temperature dependence: $n_{qp}\propto\sqrt{\Delta k_BT} \exp(-\Delta/k_B T)$. The average quasiparticle lifetime has the inverse temperature dependence\cite{skaplan1976}: $\tau_{qp}\propto\sqrt{1/\Delta k_BT} \exp(\Delta/k_B T)$. $\Delta$ is the energy gap of the superconductor, $T$ the temperature and $k_B$ Boltzmann's constant. Two quasiparticles with opposite spins and momenta can be generated from a Cooper pair by a phonon with an energy larger than $2\Delta$. When two quasiparticles recombine into a Cooper pair, a phonon is emitted. These processes are random processes in equilibrium and lead to fluctuations of the number of quasiparticles around the average. The power spectral density of these fluctuations shows a Lorentzian spectrum, given by \cite{cwilson2004}
\begin{equation}
	S_N(f)=\frac{4N_{qp}\tau_{qp}}{1+(2\pi f\tau_{qp})^2},
	\label{eq:SN}
\end{equation}
with $f$ the frequency and $N_{qp}=n_{qp}V$, with $V$ the volume of the system. Since the temperature dependences of $N_{qp}$ and $\tau_{qp}$ are exactly opposite, $N_{qp}\tau_{qp}$ is constant over temperature. This formulation was verified through fluctuations in the quasiparticle current of a Cooper pair box \cite{cwilson2001}. Recently we have observed these fluctuations in the dissipative response of a microwave resonator \cite{pdevisser2011}. As discussed above, equilibrium fluctuations in the number of quasiparticles should coincide with fluctuations in the condensate, which should show up in the reactive response of the resonator. 

%Method/Observables
To measure the complex conductivity, a 40 nm thick Al film was sputter-deposited onto a c-plane sapphire substrate. The critical temperature is 1.11 K, from which the energy gap $\Delta=1.76k_BT_c=168$ $\mu$eV. The film was patterned by wet etching into half wavelength, coplanar waveguide resonators. The resonator has a central strip volume of $1.0\cdot 10^3$ $\mu$m$^3$ and resonates at 6.62 GHz. The sample is cooled in a pulse tube pre-cooled adiabatic demagnetization refrigerator to 100 mK, with a box-in-box configuration and coax cable filters for thorough stray light shielding, crucial for these measurements. More details on the setup are given in Ref. \onlinecite{jbaselmans2012}. 

The complex transmission of the microwave circuit is measured with a quadrature mixer as a function of frequency and traces out a circle in the complex plane. The resonator amplitude, $A$, measured from the circle center, is proportional to $\sigma_1$ and therefore called the dissipation quadrature. The phase, $\theta$, is proportional to $\sigma_2$ and is also called the frequency quadrature. The responsivities of amplitude and phase to a change in the number of quasiparticles are determined experimentally as described in Ref. \onlinecite{jbaselmans2008}, which leads to $dA/dN_{qp}=-5\times 10^{-7}$ and $d\theta/dN_{qp}=4\times 10^{-6}$ at 100 mK. At the end of this Letter we will discuss the reliability of this method for a readout-power dependent quasiparticle density. The \textit{cross} power spectral density due to correlated quasiparticle number fluctuations in the resonator amplitude and phase is given by 
\begin{equation}
	S_{A,\theta}(f) = S_N(f)\frac{dA\cdotp d\theta/dN_{qp}^2}{1+(2\pi f\tau_{res})^2},
\label{eq:radiusnoise}
\end{equation}
which is only different from the amplitude or phase power spectral density by the responsivity factor, which would be $\left(dA/dN_{qp}\right)^2$ and $\left(d\theta/dN_{qp}\right)^2$ for the amplitude and phase spectra respectively. $\tau_{res}$ is the resonator ringtime given by $\tau_{res}=\frac{Q}{\pi f_0}\approx 2$ $\mu$s. Because the amplitude responsivity to quasiparticles is negative, we expect that the correlation of the quasiparticle fluctuations in amplitude and phase is negative.

%Measurement correlation
We have measured the fluctuations in the resonator amplitude and phase as a function of time at the resonant frequency. Occasionally peaks occur in the time domain data due to high energy impacts, which are filtered out of the spectral analysis as discussed in Ref. \onlinecite{pdevisser2011}. The power spectral densities of amplitude and phase are calculated by taking the Fourier transform of the autocorrelation of the time domain signals. The cross power spectral density is calculated by Fourier transforming the cross-correlation function of amplitude and phase. If the direction of the fluctuations with respect to the resonant circle in the complex plane is offset, one may convert phase noise into the amplitude direction. By simulating different orientations, we estimate the statistical error in the orientation to be $\pm 0.28^{\circ}$, which leads to an uncertainty of $\pm 0.7$ dB in the level of the cross power spectum.

\begin{figure}

\includegraphics{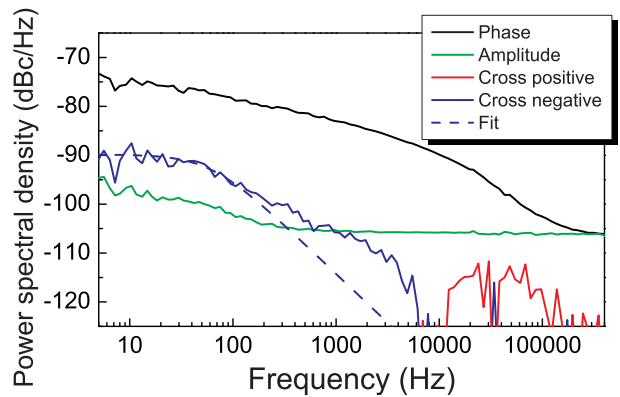} %
\caption{\label{fig:deviceandcircle} (Color online) Amplitude, phase and cross power spectral densities of the resonator as a function of frequency at 120 mK and a microwave power of $-$75 dBm. Because of the log-scale the positive and negative parts of the cross power spectral density are plotted separately. The dashed line is a single-timescale Lorentzian fit.}

\end{figure}

The amplitude, phase and cross power spectral densities at 120 mK are shown as a function of frequency in Fig. \ref{fig:deviceandcircle}. We first compare the different spectra. To start with, the phase noise is orders of magnitude higher than the amplitude noise, which is due to the response of two-level-system (TLS) fluctuators in the dielectrics to the electric field \cite{jgao2007}. Therefore observing quasiparticle fluctuations in the phase-only spectrum is nearly impossible. The flat level of $-$106 dBc/Hz is due to the amplifier noise and attenuation in between the sample and the (HEMT) amplifier and corresponds to a system noise temperature of 7 K. It was recently shown for similar microwave resonators that there is no TLS noise in the amplitude down to the vacuum noise \cite{jgao2011}. Therefore, if the system noise is subtracted, the quasiparticle signature becomes visible in the amplitude spectrum\cite{pdevisser2011}. The cross power spectrum shows no TLS noise or amplifier noise, which shows that these contributions are uncorrelated. 

We now look closer at the cross power spectral density. Part of the cross power spectrum (up to 10 kHz) is negative (blue) as expected for quasiparticle fluctuations. This part of the spectrum is real, meaning that the quasiparticle fluctuations enter amplitude and phase without relative delay. A small part at higher frequency has equal real and imaginary parts, of which the real part is positive (red). The negative part consists of two roll-offs. The first roll-off is at the quasiparticle lifetime ($\tau\approx$ 2 ms, $f\approx$ 80 Hz) as shown by the dashed line in Fig. \ref{fig:deviceandcircle}. As a function of temperature, the lifetime from the cross spectra is the same as from amplitude-only spectra (Ref. \onlinecite{pdevisser2011}), consistent with the framework of quasiparticle number fluctuations. The difference in the level of the cross and amplitude spectra ($10\pm 1$) is due to the difference in amplitude and phase responsivity. The first conclusion of this Letter is that the correlated noise in the amplitude and phase of the resonator is due to correlated fluctuations in the quasiparticle system and the superconducting condensate.

The second roll-off in Fig. \ref{fig:deviceandcircle} is at a shorter timescale ($\tau\approx$ 100 $\mu$s,  $f\approx$ 1.5 kHz) and has a much lower (a factor 25) noise level. We interpret this second roll-off in the spectrum as a signature of phonon fluctuations. This phenomenon requires a more extensive discussion which we will publish separately. The positive part of the spectrum is small and only visible close to the resonator ring-time (which also determines the roll-off frequency of the phase spectrum), which we attribute to phase-amplitude mixing due to slight detuning from the resonant frequency during the measurement \cite{jzmuidzinasbump}. The sign of this contribution varies between different resonators, where the quasiparticle contribution is always negative.

We observe that around 10 kHz the spectral density drops to below $-$120 dBc/Hz, which corresponds to the vacuum noise $\frac{1}{2}hf/k_B$. Note that the vacuum noise is not a physical limit here, but only used for comparison. Thus correlating amplitude and phase means a factor of 25 improvement with respect to the amplifier noise level when measuring amplitude or phase only, proving the high sensitivity of this method in measuring quasiparticle fluctuations. For microwave resonators used as photon detectors and limited by uncorrelated noise, we envision an improvement in sensitivity if one reads out the detector by correlating the amplitude and phase signal.

\begin{figure}

\includegraphics{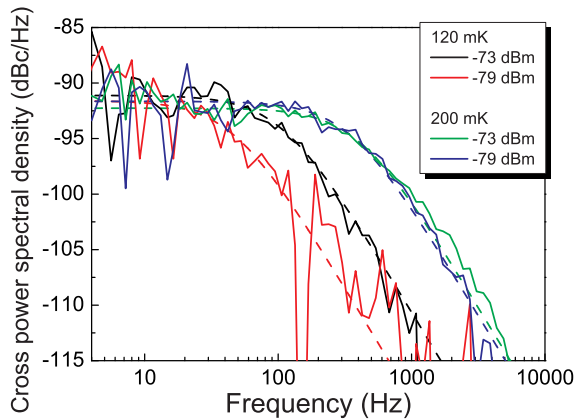}%
\caption{\label{fig:crossvsP} (Color online) Cross power spectral density of the resonator amplitude and phase as a function of frequency at two different microwave powers and temperatures of 120 and 200 mK. The dashed lines are single-timescale Lorentzian fits.}

\end{figure}

\begin{figure}
\includegraphics{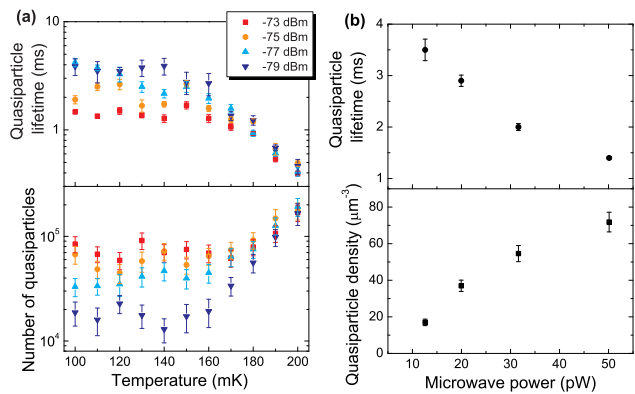} %
\caption{\label{fig:powerdependence} (Color online) (a) Number of quasiparticles and the quasiparticle lifetime as a function of temperature for four different microwave readout powers. (b) Quasiparticle density and quasiparticle lifetime as a function of readout power. Each point is a weighted average of the values from 100-150 mK as shown in (a).}

\end{figure}

%Power dependence
In Figure \ref{fig:crossvsP} cross spectra are shown at 120 mK and 200 mK for two different microwave readout powers. We observe that at 120 mK the roll-off frequency increases with increasing microwave power. This behaviour is observed up to 190 mK. At 200 mK, the spectra at different powers are similar, which marks the point where thermal quasiparticles start to dominate. We extract the number of quasiparticles from the cross power spectral density by using Eqs. \ref{eq:SN} and \ref{eq:radiusnoise} and the quasiparticle lifetimes as determined from the roll-off frequency of the spectra. The number of quasiparticles and the quasiparticle lifetime are plotted for four different readout powers as a function of temperature in Fig. \ref{fig:powerdependence}a. The error bars represent statistical errors as obtained from the fits and, for $N_{qp}$, the orientation uncertainty described above. The level at which the number of quasiparticles saturates clearly decreases with decreasing readout power. The quasiparticle lifetime increases with decreasing power, consistent with the decreasing number of quasiparticles. To get a better estimate of the saturation levels, we average $n_{qp}$ and $\tau_{qp}$ from 100-150 mK and plot the averages as a function of power in Fig. \ref{fig:powerdependence}b. We conclude that the microwave readout signal is the main source of excess quasiparticles at low temperature for superconducting resonators. 

In the broader field of superconducting quantum circuits \cite{jmartinis2009,*gcatelani2011,*hpaik2011,*mlenander2011,*rbarends2011,*acorcoles2011,osaira2012} excess quasiparticles are mainly attributed to environmental effects, which we strongly reduce by shielding our sample box and filtering the cables\cite{jbaselmans2012}. The fact that we reveal a microwave power dependence of the number of quasiparticles proves that our setup is light-tight to at least the lowest measured quasiparticle density (less than 0.1 fW of stray-light\cite{jbaselmans2012}). We note that since qubits are operated in the limit of a few microwave photons, it is unlikely that the excess quasiparticles in these systems are also due to the microwave power.

In the simplest picture, the power that is needed to create a certain number of quasiparticles is given by $P_{qp}=N_{qp}\Delta/\tau_{qp}$. Based on $P_{qp}$ and an estimate of the power that is dissipated in the quasiparticle system, $P_{diss}$, we can ascribe an efficiency $\eta_{read}$ to the process of quasiparticle creation due to the microwave readout power given by $\eta_{read}=P_{qp}/P_{diss}$. $P_{diss}$ is given by $P_{diss}=\chi_{c}\chi_{qp}P_{read}/2$, with $P_{read}$ the applied microwave power (\textit{not} the internal power in the resonator). $\chi_{c}=\frac{4Q^2}{Q_cQ_i}$ is the coupling efficiency and $\chi_{qp}=Q_i/Q_{i,qp}$ is the fraction of dissipated power that goes into the quasiparticle system. For this resonator the coupling quality factor is $Q_c=3.9\times 10^4$ and the internal quality factor is $Q_i=1.65\times 10^5$ at 100 mK and, which hardly change with power. Since we can measure the number of quasiparticles, we can estimate the quality factor due to quasiparticles, $Q_{i,qp}$, for each power, which ranges from $6\times 10^5$ at $P_{read}=-$73 dBm to $4\times 10^6$ at $P_{read}=-$79 dBm. Given these numbers we calculate $\eta_{read}=3.3\pm 1.3\times 10^{-4}$. The sensitivity of microwave resonators used as kinetic inductance detectors is usually expressed by the noise equivalent power ($NEP$). The $NEP$ due to quasiparticle number fluctuations\cite{pdevisser2012} can be expressed as $NEP=\frac{2\Delta}{\eta}\sqrt{N_{qp}/\tau_{qp}}=\frac{2}{\eta}\sqrt{\eta_{read}P_{diss}\Delta}$, with $\eta\approx 0.6$ a conversion efficiency of optical energy into quasiparticles. The measured value of $\eta_{read}$ is pleasingly low in this context. At the lowest measured readout power we get $NEP=2\times 10^{-19}$ W/Hz$^{-1/2}$.   

So far we have not touched upon the mechanism with which the readout signal leads to excess quasiparticles. We may explain the excess quasiparticles by Joule heating due to the microwave power, leading to an elevated steady state temperature of the quasiparticle system. To that end, we use the model described in Ref. \onlinecite{pdevisser2010}, in which heat transport is assumed to be limited by electron-phonon coupling. The model parameters are the same as the measured parameters of our device\footnote{We have changed the characteristic impedance equations for a microstrip geometry, used in the original model, into equations for a coplanar waveguide geometry.}. We find that $\eta_{read}$ ranges from $1\times 10^{-4}$ at the lowest readout power of $-$79 dBm to $9\times 10^{-4}$ at $-$73 dBm. Experimentally $\eta_{read}$ is constant within the uncertainty. Thus the order of magnitude of $\eta_{read}$ in the simulations agrees with the measurements, but the power dependence is different. 

The question remains how quasiparticle creation by the microwave field can be understood microscopically. A microscopic picture could be a change of the quasiparticle distribution function due to microwave absorption \cite{bivlev1973,tklapwijk1976,jchang1977} and consequently a change in the complex conductivity \cite{jchang1977, gcatelani2010} or an altered density of states due to the microwave field \cite{aanthore2003}. The redistribution of quasiparticles could lead to Cooper pair breaking. The next step to unravel the physical mechanism of the microwave power dependent quasiparticle density will therefore be a comparison of these models to resonator measurements. In this context we note that in Fig. \ref{fig:powerdependence}b, the product $N_{qp}\tau_{qp}$ is not completely constant as a function of readout power. Our analysis here is based on thermal quasiparticles, from which we derive the responsivity to quasiparticles. As the quasiparticle distribution may be non-thermal, the complex conductivity and therefore the responsivity to the number of quasiparticles may change. This could affect the derivation of $N_{qp}$ from the noise level. The measurement of $\tau_{qp}$ is fairly rigid, since it is obtained from the roll-off frequency only.

In summary, we have measured correlated fluctuations in the quasiparticle system and condensate of a superconductor, which show up in the amplitude and phase of an aluminium microwave resonator respectively. From the correlated noise spectra, we determine the number of quasiparticles and their lifetime, which both saturate at temperatures below 160 mK. The level of this saturation is microwave power dependent, showing that the microwave readout power leads to excess quasiparticles.

%Acknowledgements
We would like to thank Y.J.Y. Lankwarden for fabricating the devices. A.E. is financially supported by NWO (Veni grant 639.041.023) and JSPS Fellowship for Research Abroad.

%References
%


\begin{thebibliography}{28}%
\makeatletter
\providecommand \@ifxundefined [1]{%
 \@ifx{#1\undefined}
}%
\providecommand \@ifnum [1]{%
 \ifnum #1\expandafter \@firstoftwo
 \else \expandafter \@secondoftwo
 \fi
}%
\providecommand \@ifx [1]{%
 \ifx #1\expandafter \@firstoftwo
 \else \expandafter \@secondoftwo
 \fi
}%
\providecommand \natexlab [1]{#1}%
\providecommand \enquote  [1]{``#1''}%
\providecommand \bibnamefont  [1]{#1}%
\providecommand \bibfnamefont [1]{#1}%
\providecommand \citenamefont [1]{#1}%
\providecommand \href@noop [0]{\@secondoftwo}%
\providecommand \href [0]{\begingroup \@sanitize@url \@href}%
\providecommand \@href[1]{\@@startlink{#1}\@@href}%
\providecommand \@@href[1]{\endgroup#1\@@endlink}%
\providecommand \@sanitize@url [0]{\catcode `\\12\catcode `\$12\catcode
  `\&12\catcode `\#12\catcode `\^12\catcode `\_12\catcode `\%12\relax}%
\providecommand \@@startlink[1]{}%
\providecommand \@@endlink[0]{}%
\providecommand \url  [0]{\begingroup\@sanitize@url \@url }%
\providecommand \@url [1]{\endgroup\@href {#1}{\urlprefix }}%
\providecommand \urlprefix  [0]{URL }%
\providecommand \Eprint [0]{\href }%
\providecommand \doibase [0]{http://dx.doi.org/}%
\providecommand \selectlanguage [0]{\@gobble}%
\providecommand \bibinfo  [0]{\@secondoftwo}%
\providecommand \bibfield  [0]{\@secondoftwo}%
\providecommand \translation [1]{[#1]}%
\providecommand \BibitemOpen [0]{}%
\providecommand \bibitemStop [0]{}%
\providecommand \bibitemNoStop [0]{.\EOS\space}%
\providecommand \EOS [0]{\spacefactor3000\relax}%
\providecommand \BibitemShut  [1]{\csname bibitem#1\endcsname}%
\let\auto@bib@innerbib\@empty
%</preamble>
\bibitem [{\citenamefont {Martinis}, \citenamefont {Ansmann},\ and\
  \citenamefont {Aumentado}(2009)}]{jmartinis2009}%
  \BibitemOpen
  \bibfield  {author} {\bibinfo {author} {\bibfnamefont {J.~M.}\ \bibnamefont
  {Martinis}}, \bibinfo {author} {\bibfnamefont {M.}~\bibnamefont {Ansmann}}, \
  and\ \bibinfo {author} {\bibfnamefont {J.}~\bibnamefont {Aumentado}},\
  }\href@noop {} {\bibfield  {journal} {\bibinfo  {journal} {Phys. Rev. Lett.}\
  }\textbf {\bibinfo {volume} {103}},\ \bibinfo {pages} {097002} (\bibinfo
  {year} {2009})}\BibitemShut {NoStop}%
\bibitem [{\citenamefont {Catelani}\ \emph {et~al.}(2011)\citenamefont
  {Catelani}, \citenamefont {Koch}, \citenamefont {Frunzio}, \citenamefont
  {Schoelkopf}, \citenamefont {Devoret},\ and\ \citenamefont
  {Glazman}}]{gcatelani2011}%
  \BibitemOpen
  \bibfield  {author} {\bibinfo {author} {\bibfnamefont {G.}~\bibnamefont
  {Catelani}}, \bibinfo {author} {\bibfnamefont {J.}~\bibnamefont {Koch}},
  \bibinfo {author} {\bibfnamefont {L.}~\bibnamefont {Frunzio}}, \bibinfo
  {author} {\bibfnamefont {R.~J.}\ \bibnamefont {Schoelkopf}}, \bibinfo
  {author} {\bibfnamefont {M.~H.}\ \bibnamefont {Devoret}}, \ and\ \bibinfo
  {author} {\bibfnamefont {L.~I.}\ \bibnamefont {Glazman}},\ }\href@noop {}
  {\bibfield  {journal} {\bibinfo  {journal} {Phys. Rev. Lett.}\ }\textbf
  {\bibinfo {volume} {106}},\ \bibinfo {pages} {077002} (\bibinfo {year}
  {2011})}\BibitemShut {NoStop}%
\bibitem [{\citenamefont {Paik}\ \emph {et~al.}(2011)\citenamefont {Paik},
  \citenamefont {Schuster}, \citenamefont {Bishop}, \citenamefont {Kirchmair},
  \citenamefont {Catelani}, \citenamefont {Sears}, \citenamefont {Johnson},
  \citenamefont {Reagor}, \citenamefont {Frunzio}, \citenamefont {Glazman},\
  and\ \citenamefont {Schoelkopf}}]{hpaik2011}%
  \BibitemOpen
  \bibfield  {author} {\bibinfo {author} {\bibfnamefont {H.}~\bibnamefont
  {Paik}}, \bibinfo {author} {\bibfnamefont {D.~I.}\ \bibnamefont {Schuster}},
  \bibinfo {author} {\bibfnamefont {L.~S.}\ \bibnamefont {Bishop}}, \bibinfo
  {author} {\bibfnamefont {G.}~\bibnamefont {Kirchmair}}, \bibinfo {author}
  {\bibfnamefont {G.}~\bibnamefont {Catelani}}, \bibinfo {author}
  {\bibfnamefont {A.~P.}\ \bibnamefont {Sears}}, \bibinfo {author}
  {\bibfnamefont {B.~R.}\ \bibnamefont {Johnson}}, \bibinfo {author}
  {\bibfnamefont {M.~J.}\ \bibnamefont {Reagor}}, \bibinfo {author}
  {\bibfnamefont {L.}~\bibnamefont {Frunzio}}, \bibinfo {author} {\bibfnamefont
  {L.}~\bibnamefont {Glazman}}, \ and\ \bibinfo {author} {\bibfnamefont
  {R.~J.}\ \bibnamefont {Schoelkopf}},\ }\href@noop {} {\bibfield  {journal}
  {\bibinfo  {journal} {Phys. Rev. Lett.}\ }\textbf {\bibinfo {volume} {107}},\
  \bibinfo {pages} {240501} (\bibinfo {year} {2011})}\BibitemShut {NoStop}%
\bibitem [{\citenamefont {Lenander}\ \emph {et~al.}(2011)\citenamefont
  {Lenander}, \citenamefont {Wang}, \citenamefont {Bialczak}, \citenamefont
  {Lucero}, \citenamefont {Mariantoni}, \citenamefont {Neeley}, \citenamefont
  {O'Connell}, \citenamefont {Sank}, \citenamefont {Weides}, \citenamefont
  {Wenner}, \citenamefont {Yamamoto}, \citenamefont {Yin}, \citenamefont
  {Zhao}, \citenamefont {Cleland},\ and\ \citenamefont
  {Martinis}}]{mlenander2011}%
  \BibitemOpen
  \bibfield  {author} {\bibinfo {author} {\bibfnamefont {M.}~\bibnamefont
  {Lenander}}, \bibinfo {author} {\bibfnamefont {H.}~\bibnamefont {Wang}},
  \bibinfo {author} {\bibfnamefont {R.~C.}\ \bibnamefont {Bialczak}}, \bibinfo
  {author} {\bibfnamefont {E.}~\bibnamefont {Lucero}}, \bibinfo {author}
  {\bibfnamefont {M.}~\bibnamefont {Mariantoni}}, \bibinfo {author}
  {\bibfnamefont {M.}~\bibnamefont {Neeley}}, \bibinfo {author} {\bibfnamefont
  {A.~D.}\ \bibnamefont {O'Connell}}, \bibinfo {author} {\bibfnamefont
  {D.}~\bibnamefont {Sank}}, \bibinfo {author} {\bibfnamefont {M.}~\bibnamefont
  {Weides}}, \bibinfo {author} {\bibfnamefont {J.}~\bibnamefont {Wenner}},
  \bibinfo {author} {\bibfnamefont {T.}~\bibnamefont {Yamamoto}}, \bibinfo
  {author} {\bibfnamefont {Y.}~\bibnamefont {Yin}}, \bibinfo {author}
  {\bibfnamefont {J.}~\bibnamefont {Zhao}}, \bibinfo {author} {\bibfnamefont
  {A.~N.}\ \bibnamefont {Cleland}}, \ and\ \bibinfo {author} {\bibfnamefont
  {J.~M.}\ \bibnamefont {Martinis}},\ }\href@noop {} {\bibfield  {journal}
  {\bibinfo  {journal} {Phys. Rev. B}\ }\textbf {\bibinfo {volume} {84}},\
  \bibinfo {pages} {024501} (\bibinfo {year} {2011})}\BibitemShut {NoStop}%
\bibitem [{\citenamefont {Barends}\ \emph {et~al.}(2011)\citenamefont
  {Barends}, \citenamefont {Wenner}, \citenamefont {Lenander}, \citenamefont
  {Chen}, \citenamefont {Bialczak}, \citenamefont {Kelly}, \citenamefont
  {Lucero}, \citenamefont {O'Malley}, \citenamefont {Mariantoni}, \citenamefont
  {Sank}, \citenamefont {Wang}, \citenamefont {White}, \citenamefont {Yin},
  \citenamefont {Zhao}, \citenamefont {Cleland}, \citenamefont {Martinis},\
  and\ \citenamefont {Baselmans}}]{rbarends2011}%
  \BibitemOpen
  \bibfield  {author} {\bibinfo {author} {\bibfnamefont {R.}~\bibnamefont
  {Barends}}, \bibinfo {author} {\bibfnamefont {J.}~\bibnamefont {Wenner}},
  \bibinfo {author} {\bibfnamefont {M.}~\bibnamefont {Lenander}}, \bibinfo
  {author} {\bibfnamefont {Y.}~\bibnamefont {Chen}}, \bibinfo {author}
  {\bibfnamefont {R.~C.}\ \bibnamefont {Bialczak}}, \bibinfo {author}
  {\bibfnamefont {J.}~\bibnamefont {Kelly}}, \bibinfo {author} {\bibfnamefont
  {E.}~\bibnamefont {Lucero}}, \bibinfo {author} {\bibfnamefont
  {P.}~\bibnamefont {O'Malley}}, \bibinfo {author} {\bibfnamefont
  {M.}~\bibnamefont {Mariantoni}}, \bibinfo {author} {\bibfnamefont
  {D.}~\bibnamefont {Sank}}, \bibinfo {author} {\bibfnamefont {H.}~\bibnamefont
  {Wang}}, \bibinfo {author} {\bibfnamefont {T.~C.}\ \bibnamefont {White}},
  \bibinfo {author} {\bibfnamefont {Y.}~\bibnamefont {Yin}}, \bibinfo {author}
  {\bibfnamefont {J.}~\bibnamefont {Zhao}}, \bibinfo {author} {\bibfnamefont
  {A.~N.}\ \bibnamefont {Cleland}}, \bibinfo {author} {\bibfnamefont {J.~M.}\
  \bibnamefont {Martinis}}, \ and\ \bibinfo {author} {\bibfnamefont {J.~J.~A.}\
  \bibnamefont {Baselmans}},\ }\href@noop {} {\bibfield  {journal} {\bibinfo
  {journal} {Appl. Phys. Lett.}\ }\textbf {\bibinfo {volume} {99}},\ \bibinfo
  {pages} {113507} (\bibinfo {year} {2011})}\BibitemShut {NoStop}%
\bibitem [{\citenamefont {C\'{o}rcoles}\ \emph {et~al.}(2011)\citenamefont
  {C\'{o}rcoles}, \citenamefont {Chow}, \citenamefont {Gambetta}, \citenamefont
  {Rigetti}, \citenamefont {Rozen}, \citenamefont {Keefe}, \citenamefont
  {Rothwell}, \citenamefont {Ketchen},\ and\ \citenamefont
  {Steffen}}]{acorcoles2011}%
  \BibitemOpen
  \bibfield  {author} {\bibinfo {author} {\bibfnamefont {A.~D.}\ \bibnamefont
  {C\'{o}rcoles}}, \bibinfo {author} {\bibfnamefont {J.~M.}\ \bibnamefont
  {Chow}}, \bibinfo {author} {\bibfnamefont {J.~M.}\ \bibnamefont {Gambetta}},
  \bibinfo {author} {\bibfnamefont {C.}~\bibnamefont {Rigetti}}, \bibinfo
  {author} {\bibfnamefont {J.~R.}\ \bibnamefont {Rozen}}, \bibinfo {author}
  {\bibfnamefont {G.~A.}\ \bibnamefont {Keefe}}, \bibinfo {author}
  {\bibfnamefont {M.~B.}\ \bibnamefont {Rothwell}}, \bibinfo {author}
  {\bibfnamefont {M.~B.}\ \bibnamefont {Ketchen}}, \ and\ \bibinfo {author}
  {\bibfnamefont {M.}~\bibnamefont {Steffen}},\ }\href@noop {} {\bibfield
  {journal} {\bibinfo  {journal} {Appl. Phys. Lett.}\ }\textbf {\bibinfo
  {volume} {99}},\ \bibinfo {pages} {181906} (\bibinfo {year}
  {2011})}\BibitemShut {NoStop}%
\bibitem [{\citenamefont {Saira}\ \emph {et~al.}(2012)\citenamefont {Saira},
  \citenamefont {Kemppinen}, \citenamefont {Maisi},\ and\ \citenamefont
  {Pekola}}]{osaira2012}%
  \BibitemOpen
  \bibfield  {author} {\bibinfo {author} {\bibfnamefont {O.-P.}\ \bibnamefont
  {Saira}}, \bibinfo {author} {\bibfnamefont {A.}~\bibnamefont {Kemppinen}},
  \bibinfo {author} {\bibfnamefont {V.~F.}\ \bibnamefont {Maisi}}, \ and\
  \bibinfo {author} {\bibfnamefont {J.~P.}\ \bibnamefont {Pekola}},\
  }\href@noop {} {\bibfield  {journal} {\bibinfo  {journal} {Phys. Rev. B}\
  }\textbf {\bibinfo {volume} {85}},\ \bibinfo {pages} {012504} (\bibinfo
  {year} {2012})}\BibitemShut {NoStop}%
\bibitem [{\citenamefont {Barends}\ \emph {et~al.}(2008)\citenamefont
  {Barends}, \citenamefont {Baselmans}, \citenamefont {Yates}, \citenamefont
  {Gao}, \citenamefont {Hovenier},\ and\ \citenamefont
  {Klapwijk}}]{rbarends2008c}%
  \BibitemOpen
  \bibfield  {author} {\bibinfo {author} {\bibfnamefont {R.}~\bibnamefont
  {Barends}}, \bibinfo {author} {\bibfnamefont {J.~J.~A.}\ \bibnamefont
  {Baselmans}}, \bibinfo {author} {\bibfnamefont {S.~J.~C.}\ \bibnamefont
  {Yates}}, \bibinfo {author} {\bibfnamefont {J.~R.}\ \bibnamefont {Gao}},
  \bibinfo {author} {\bibfnamefont {J.~N.}\ \bibnamefont {Hovenier}}, \ and\
  \bibinfo {author} {\bibfnamefont {T.~M.}\ \bibnamefont {Klapwijk}},\
  }\href@noop {} {\bibfield  {journal} {\bibinfo  {journal} {Phys. Rev. Lett.}\
  }\textbf {\bibinfo {volume} {100}},\ \bibinfo {pages} {257002} (\bibinfo
  {year} {2008})}\BibitemShut {NoStop}%
\bibitem [{\citenamefont {de~Visser}\ \emph {et~al.}(2011)\citenamefont
  {de~Visser}, \citenamefont {Baselmans}, \citenamefont {Diener}, \citenamefont
  {Yates}, \citenamefont {Endo},\ and\ \citenamefont
  {Klapwijk}}]{pdevisser2011}%
  \BibitemOpen
  \bibfield  {author} {\bibinfo {author} {\bibfnamefont {P.~J.}\ \bibnamefont
  {de~Visser}}, \bibinfo {author} {\bibfnamefont {J.~J.~A.}\ \bibnamefont
  {Baselmans}}, \bibinfo {author} {\bibfnamefont {P.}~\bibnamefont {Diener}},
  \bibinfo {author} {\bibfnamefont {S.~J.~C.}\ \bibnamefont {Yates}}, \bibinfo
  {author} {\bibfnamefont {A.}~\bibnamefont {Endo}}, \ and\ \bibinfo {author}
  {\bibfnamefont {T.~M.}\ \bibnamefont {Klapwijk}},\ }\href@noop {} {\bibfield
  {journal} {\bibinfo  {journal} {Phys. Rev. Lett.}\ }\textbf {\bibinfo
  {volume} {106}},\ \bibinfo {pages} {167004} (\bibinfo {year}
  {2011})}\BibitemShut {NoStop}%
\bibitem [{\citenamefont {Day}\ \emph {et~al.}(2003)\citenamefont {Day},
  \citenamefont {LeDuc}, \citenamefont {Mazin}, \citenamefont {Vayonakis},\
  and\ \citenamefont {Zmuidzinas}}]{pday2003}%
  \BibitemOpen
  \bibfield  {author} {\bibinfo {author} {\bibfnamefont {P.~K.}\ \bibnamefont
  {Day}}, \bibinfo {author} {\bibfnamefont {H.~G.}\ \bibnamefont {LeDuc}},
  \bibinfo {author} {\bibfnamefont {B.~A.}\ \bibnamefont {Mazin}}, \bibinfo
  {author} {\bibfnamefont {A.}~\bibnamefont {Vayonakis}}, \ and\ \bibinfo
  {author} {\bibfnamefont {J.}~\bibnamefont {Zmuidzinas}},\ }\href@noop {}
  {\bibfield  {journal} {\bibinfo  {journal} {Nature}\ }\textbf {\bibinfo
  {volume} {425}},\ \bibinfo {pages} {817} (\bibinfo {year}
  {2003})}\BibitemShut {NoStop}%
\bibitem [{\citenamefont {Wallraff}\ \emph {et~al.}(2004)\citenamefont
  {Wallraff}, \citenamefont {Schuster}, \citenamefont {Blais}, \citenamefont
  {Frunzio}, \citenamefont {Huang}, \citenamefont {Majer}, \citenamefont
  {Kumar}, \citenamefont {Girvin},\ and\ \citenamefont
  {Schoelkopf}}]{awallraff2004}%
  \BibitemOpen
  \bibfield  {author} {\bibinfo {author} {\bibfnamefont {A.}~\bibnamefont
  {Wallraff}}, \bibinfo {author} {\bibfnamefont {D.~I.}\ \bibnamefont
  {Schuster}}, \bibinfo {author} {\bibfnamefont {A.}~\bibnamefont {Blais}},
  \bibinfo {author} {\bibfnamefont {L.}~\bibnamefont {Frunzio}}, \bibinfo
  {author} {\bibfnamefont {R.-S.}\ \bibnamefont {Huang}}, \bibinfo {author}
  {\bibfnamefont {J.}~\bibnamefont {Majer}}, \bibinfo {author} {\bibfnamefont
  {S.}~\bibnamefont {Kumar}}, \bibinfo {author} {\bibfnamefont {S.~M.}\
  \bibnamefont {Girvin}}, \ and\ \bibinfo {author} {\bibfnamefont {R.~J.}\
  \bibnamefont {Schoelkopf}},\ }\href@noop {} {\bibfield  {journal} {\bibinfo
  {journal} {Nature}\ }\textbf {\bibinfo {volume} {431}},\ \bibinfo {pages}
  {162} (\bibinfo {year} {2004})}\BibitemShut {NoStop}%
\bibitem [{\citenamefont {Mattis}\ and\ \citenamefont
  {Bardeen}(1958)}]{dmattis1958}%
  \BibitemOpen
  \bibfield  {author} {\bibinfo {author} {\bibfnamefont {D.~C.}\ \bibnamefont
  {Mattis}}\ and\ \bibinfo {author} {\bibfnamefont {J.}~\bibnamefont
  {Bardeen}},\ }\href@noop {} {\bibfield  {journal} {\bibinfo  {journal} {Phys.
  Rev.}\ }\textbf {\bibinfo {volume} {111}},\ \bibinfo {pages} {412} (\bibinfo
  {year} {1958})}\BibitemShut {NoStop}%
\bibitem [{\citenamefont {Gao}\ \emph {et~al.}(2007)\citenamefont {Gao},
  \citenamefont {Zmuidzinas}, \citenamefont {Mazin}, \citenamefont {LeDuc},\
  and\ \citenamefont {Day}}]{jgao2007}%
  \BibitemOpen
  \bibfield  {author} {\bibinfo {author} {\bibfnamefont {J.}~\bibnamefont
  {Gao}}, \bibinfo {author} {\bibfnamefont {J.}~\bibnamefont {Zmuidzinas}},
  \bibinfo {author} {\bibfnamefont {B.~A.}\ \bibnamefont {Mazin}}, \bibinfo
  {author} {\bibfnamefont {H.~G.}\ \bibnamefont {LeDuc}}, \ and\ \bibinfo
  {author} {\bibfnamefont {P.~K.}\ \bibnamefont {Day}},\ }\href@noop {}
  {\bibfield  {journal} {\bibinfo  {journal} {Appl. Phys. Lett.}\ }\textbf
  {\bibinfo {volume} {90}},\ \bibinfo {pages} {102507} (\bibinfo {year}
  {2007})}\BibitemShut {NoStop}%
\bibitem [{\citenamefont {Kaplan}\ \emph {et~al.}(1976)\citenamefont {Kaplan},
  \citenamefont {Chi}, \citenamefont {Langenberg}, \citenamefont {Chang},
  \citenamefont {Jafarey},\ and\ \citenamefont {Scalapino}}]{skaplan1976}%
  \BibitemOpen
  \bibfield  {author} {\bibinfo {author} {\bibfnamefont {S.~B.}\ \bibnamefont
  {Kaplan}}, \bibinfo {author} {\bibfnamefont {C.~C.}\ \bibnamefont {Chi}},
  \bibinfo {author} {\bibfnamefont {D.~N.}\ \bibnamefont {Langenberg}},
  \bibinfo {author} {\bibfnamefont {J.}~\bibnamefont {Chang}}, \bibinfo
  {author} {\bibfnamefont {S.}~\bibnamefont {Jafarey}}, \ and\ \bibinfo
  {author} {\bibfnamefont {D.~J.}\ \bibnamefont {Scalapino}},\ }\href@noop {}
  {\bibfield  {journal} {\bibinfo  {journal} {Phys. Rev. B}\ }\textbf {\bibinfo
  {volume} {14}},\ \bibinfo {pages} {11} (\bibinfo {year} {1976})}\BibitemShut
  {NoStop}%
\bibitem [{\citenamefont {Wilson}\ and\ \citenamefont
  {Prober}(2004)}]{cwilson2004}%
  \BibitemOpen
  \bibfield  {author} {\bibinfo {author} {\bibfnamefont {C.~M.}\ \bibnamefont
  {Wilson}}\ and\ \bibinfo {author} {\bibfnamefont {D.~E.}\ \bibnamefont
  {Prober}},\ }\href@noop {} {\bibfield  {journal} {\bibinfo  {journal} {Phys.
  Rev. B}\ }\textbf {\bibinfo {volume} {69}},\ \bibinfo {pages} {094524}
  (\bibinfo {year} {2004})}\BibitemShut {NoStop}%
\bibitem [{\citenamefont {Wilson}, \citenamefont {Frunzio},\ and\ \citenamefont
  {Prober}(2001)}]{cwilson2001}%
  \BibitemOpen
  \bibfield  {author} {\bibinfo {author} {\bibfnamefont {C.~M.}\ \bibnamefont
  {Wilson}}, \bibinfo {author} {\bibfnamefont {L.}~\bibnamefont {Frunzio}}, \
  and\ \bibinfo {author} {\bibfnamefont {D.~E.}\ \bibnamefont {Prober}},\
  }\href@noop {} {\bibfield  {journal} {\bibinfo  {journal} {Phys. Rev. Lett.}\
  }\textbf {\bibinfo {volume} {87}},\ \bibinfo {pages} {067004} (\bibinfo
  {year} {2001})}\BibitemShut {NoStop}%
\bibitem [{\citenamefont {Baselmans}\ \emph {et~al.}(2012)\citenamefont
  {Baselmans}, \citenamefont {Yates}, \citenamefont {Diener},\ and\
  \citenamefont {de~Visser}}]{jbaselmans2012}%
  \BibitemOpen
  \bibfield  {author} {\bibinfo {author} {\bibfnamefont {J.}~\bibnamefont
  {Baselmans}}, \bibinfo {author} {\bibfnamefont {S.}~\bibnamefont {Yates}},
  \bibinfo {author} {\bibfnamefont {P.}~\bibnamefont {Diener}}, \ and\ \bibinfo
  {author} {\bibfnamefont {P.}~\bibnamefont {de~Visser}},\ }\href@noop {}
  {\bibfield  {journal} {\bibinfo  {journal} {J. Low Temp. Phys.}\ }\textbf
  {\bibinfo {volume} {167}},\ \bibinfo {pages} {360} (\bibinfo {year}
  {2012})}\BibitemShut {NoStop}%
\bibitem [{\citenamefont {Baselmans}\ \emph {et~al.}(2008)\citenamefont
  {Baselmans}, \citenamefont {Yates}, \citenamefont {Barends}, \citenamefont
  {Lankwarden}, \citenamefont {Gao}, \citenamefont {Hoevers},\ and\
  \citenamefont {Klapwijk}}]{jbaselmans2008}%
  \BibitemOpen
  \bibfield  {author} {\bibinfo {author} {\bibfnamefont {J.}~\bibnamefont
  {Baselmans}}, \bibinfo {author} {\bibfnamefont {S.~J.~C.}\ \bibnamefont
  {Yates}}, \bibinfo {author} {\bibfnamefont {R.}~\bibnamefont {Barends}},
  \bibinfo {author} {\bibfnamefont {Y.~J.~Y.}\ \bibnamefont {Lankwarden}},
  \bibinfo {author} {\bibfnamefont {J.~R.}\ \bibnamefont {Gao}}, \bibinfo
  {author} {\bibfnamefont {H.}~\bibnamefont {Hoevers}}, \ and\ \bibinfo
  {author} {\bibfnamefont {T.~M.}\ \bibnamefont {Klapwijk}},\ }\href@noop {}
  {\bibfield  {journal} {\bibinfo  {journal} {J. Low Temp. Phys.}\ }\textbf
  {\bibinfo {volume} {151}},\ \bibinfo {pages} {524} (\bibinfo {year}
  {2008})}\BibitemShut {NoStop}%
\bibitem [{\citenamefont {Gao}\ \emph {et~al.}(2011)\citenamefont {Gao},
  \citenamefont {Vale}, \citenamefont {Mates}, \citenamefont {Schmidt},
  \citenamefont {Hilton}, \citenamefont {Irwin}, \citenamefont {Mallet},
  \citenamefont {Castellanos-Beltran}, \citenamefont {Lehnert}, \citenamefont
  {Zmuidzinas},\ and\ \citenamefont {LeDuc}}]{jgao2011}%
  \BibitemOpen
  \bibfield  {author} {\bibinfo {author} {\bibfnamefont {J.}~\bibnamefont
  {Gao}}, \bibinfo {author} {\bibfnamefont {L.~R.}\ \bibnamefont {Vale}},
  \bibinfo {author} {\bibfnamefont {J.~A.~B.}\ \bibnamefont {Mates}}, \bibinfo
  {author} {\bibfnamefont {D.~R.}\ \bibnamefont {Schmidt}}, \bibinfo {author}
  {\bibfnamefont {G.~C.}\ \bibnamefont {Hilton}}, \bibinfo {author}
  {\bibfnamefont {K.~D.}\ \bibnamefont {Irwin}}, \bibinfo {author}
  {\bibfnamefont {F.}~\bibnamefont {Mallet}}, \bibinfo {author} {\bibfnamefont
  {M.}~\bibnamefont {Castellanos-Beltran}}, \bibinfo {author} {\bibfnamefont
  {K.~W.}\ \bibnamefont {Lehnert}}, \bibinfo {author} {\bibfnamefont
  {J.}~\bibnamefont {Zmuidzinas}}, \ and\ \bibinfo {author} {\bibfnamefont
  {H.~G.}\ \bibnamefont {LeDuc}},\ }\href@noop {} {\bibfield  {journal}
  {\bibinfo  {journal} {Appl. Phys. Lett.}\ }\textbf {\bibinfo {volume} {98}},\
  \bibinfo {pages} {232508} (\bibinfo {year} {2011})}\BibitemShut {NoStop}%
\bibitem [{\citenamefont {Zmuidzinas}\ \emph {et~al.}(shed)\citenamefont
  {Zmuidzinas}, \citenamefont {Gao}, \citenamefont {Day},\ and\ \citenamefont
  {LeDuc}}]{jzmuidzinasbump}%
  \BibitemOpen
  \bibfield  {author} {\bibinfo {author} {\bibfnamefont {J.}~\bibnamefont
  {Zmuidzinas}}, \bibinfo {author} {\bibfnamefont {J.}~\bibnamefont {Gao}},
  \bibinfo {author} {\bibfnamefont {P.~K.}\ \bibnamefont {Day}}, \ and\
  \bibinfo {author} {\bibfnamefont {H.~G.}\ \bibnamefont {LeDuc}},\ }\href@noop
  {} {} (\bibinfo {year} {unpublished})\BibitemShut {NoStop}%
\bibitem [{\citenamefont {de~Visser}\ \emph {et~al.}(2012)\citenamefont
  {de~Visser}, \citenamefont {Baselmans}, \citenamefont {Diener}, \citenamefont
  {Yates}, \citenamefont {Endo},\ and\ \citenamefont
  {Klapwijk}}]{pdevisser2012}%
  \BibitemOpen
  \bibfield  {author} {\bibinfo {author} {\bibfnamefont {P.~J.}\ \bibnamefont
  {de~Visser}}, \bibinfo {author} {\bibfnamefont {J.~J.~A.}\ \bibnamefont
  {Baselmans}}, \bibinfo {author} {\bibfnamefont {P.}~\bibnamefont {Diener}},
  \bibinfo {author} {\bibfnamefont {S.~J.~C.}\ \bibnamefont {Yates}}, \bibinfo
  {author} {\bibfnamefont {A.}~\bibnamefont {Endo}}, \ and\ \bibinfo {author}
  {\bibfnamefont {T.~M.}\ \bibnamefont {Klapwijk}},\ }\href@noop {} {\bibfield
  {journal} {\bibinfo  {journal} {J. Low Temp. Phys.}\ }\textbf {\bibinfo
  {volume} {167}},\ \bibinfo {pages} {335} (\bibinfo {year}
  {2012})}\BibitemShut {NoStop}%
\bibitem [{\citenamefont {de~Visser}, \citenamefont {Withington},\ and\
  \citenamefont {Goldie}(2010)}]{pdevisser2010}%
  \BibitemOpen
  \bibfield  {author} {\bibinfo {author} {\bibfnamefont {P.~J.}\ \bibnamefont
  {de~Visser}}, \bibinfo {author} {\bibfnamefont {S.}~\bibnamefont
  {Withington}}, \ and\ \bibinfo {author} {\bibfnamefont {D.~J.}\ \bibnamefont
  {Goldie}},\ }\href@noop {} {\bibfield  {journal} {\bibinfo  {journal} {J.
  Appl. Phys.}\ }\textbf {\bibinfo {volume} {108}},\ \bibinfo {pages} {114504}
  (\bibinfo {year} {2010})}\BibitemShut {NoStop}%
\bibitem [{Note1()}]{Note1}%
  \BibitemOpen
  \bibinfo {note} {We have changed the characteristic impedance equations for a
  microstrip geometry, used in the original model, into equations for a
  coplanar waveguide geometry.}\BibitemShut {Stop}%
\bibitem [{\citenamefont {Ivlev}, \citenamefont {Lisitsyn},\ and\ \citenamefont
  {Eliashberg}(1973)}]{bivlev1973}%
  \BibitemOpen
  \bibfield  {author} {\bibinfo {author} {\bibfnamefont {B.~I.}\ \bibnamefont
  {Ivlev}}, \bibinfo {author} {\bibfnamefont {S.~G.}\ \bibnamefont {Lisitsyn}},
  \ and\ \bibinfo {author} {\bibfnamefont {G.~M.}\ \bibnamefont {Eliashberg}},\
  }\href@noop {} {\bibfield  {journal} {\bibinfo  {journal} {J. Low Temp.
  Phys.}\ }\textbf {\bibinfo {volume} {10}},\ \bibinfo {pages} {449} (\bibinfo
  {year} {1973})}\BibitemShut {NoStop}%
\bibitem [{\citenamefont {Klapwijk}\ and\ \citenamefont
  {Mooij}(1976)}]{tklapwijk1976}%
  \BibitemOpen
  \bibfield  {author} {\bibinfo {author} {\bibfnamefont {T.~M.}\ \bibnamefont
  {Klapwijk}}\ and\ \bibinfo {author} {\bibfnamefont {J.~E.}\ \bibnamefont
  {Mooij}},\ }\href@noop {} {\bibfield  {journal} {\bibinfo  {journal} {Physica
  B}\ }\textbf {\bibinfo {volume} {81}},\ \bibinfo {pages} {132} (\bibinfo
  {year} {1976})}\BibitemShut {NoStop}%
\bibitem [{\citenamefont {Chang}\ and\ \citenamefont
  {Scalapino}(1977)}]{jchang1977}%
  \BibitemOpen
  \bibfield  {author} {\bibinfo {author} {\bibfnamefont {J.-J.}\ \bibnamefont
  {Chang}}\ and\ \bibinfo {author} {\bibfnamefont {D.~J.}\ \bibnamefont
  {Scalapino}},\ }\href@noop {} {\bibfield  {journal} {\bibinfo  {journal}
  {Phys. Rev. B}\ }\textbf {\bibinfo {volume} {15}},\ \bibinfo {pages} {2651}
  (\bibinfo {year} {1977})}\BibitemShut {NoStop}%
\bibitem [{\citenamefont {Catelani}, \citenamefont {Glazman},\ and\
  \citenamefont {Nagaev}(2010)}]{gcatelani2010}%
  \BibitemOpen
  \bibfield  {author} {\bibinfo {author} {\bibfnamefont {G.}~\bibnamefont
  {Catelani}}, \bibinfo {author} {\bibfnamefont {L.~I.}\ \bibnamefont
  {Glazman}}, \ and\ \bibinfo {author} {\bibfnamefont {K.~E.}\ \bibnamefont
  {Nagaev}},\ }\href@noop {} {\bibfield  {journal} {\bibinfo  {journal} {Phys.
  Rev. B}\ }\textbf {\bibinfo {volume} {82}},\ \bibinfo {pages} {134502}
  (\bibinfo {year} {2010})}\BibitemShut {NoStop}%
\bibitem [{\citenamefont {Anthore}, \citenamefont {Pothier},\ and\
  \citenamefont {Esteve}(2003)}]{aanthore2003}%
  \BibitemOpen
  \bibfield  {author} {\bibinfo {author} {\bibfnamefont {A.}~\bibnamefont
  {Anthore}}, \bibinfo {author} {\bibfnamefont {H.}~\bibnamefont {Pothier}}, \
  and\ \bibinfo {author} {\bibfnamefont {D.}~\bibnamefont {Esteve}},\
  }\href@noop {} {\bibfield  {journal} {\bibinfo  {journal} {Phys. Rev. Lett.}\
  }\textbf {\bibinfo {volume} {90}},\ \bibinfo {pages} {127001} (\bibinfo
  {year} {2003})}\BibitemShut {NoStop}%
\end{thebibliography}
\end{document}